\documentclass[aps,pre,reprint]{revtex4-2}

\usepackage{graphicx}
\usepackage{amsmath}
\usepackage{natbib}
\usepackage{hyperref}
\hypersetup{
    colorlinks = true,
    urlcolor   = blue,
    citecolor  = blue,
}

\newcommand{\nn}{\nonumber}
\newcommand{\veps}{\varepsilon}
\newcommand{\tl}{\tilde}
\newcommand{\bea}{\begin{eqnarray}}
\newcommand{\eea}{\end{eqnarray}}

\begin{document}

\title{Electrolubrication in liquid mixtures between two parallel plates}

\author{Roni Kroll}
\email[]{krollr@post.bgu.ac.il}
\author{Yoav Tsori}
\email[]{tsori@bgu.ac.il}

\affiliation{Department of Chemical Engineering, Ben-Gurion University of the Negev, 
Rager street, Beer-Sheva 84105, Israel}

\begin{abstract}

We describe theoretically ``electrolubrication'' in liquid mixtures, the phenomenon where an electric 
field applied transverse to the confining surfaces leads to concentration gradients that alter the flow 
profile significantly. When the more polar liquid is the less viscous one, the stress in the liquid falls on 
two electric-field-induced thin lubrication layers. The thickness of the lubrication layer depends on the 
Debye length and the mixture correlation length. For the simple case of two parallel and infinite plates, 
we calculate explicitly the liquid velocity profile and integrated flux. The maximum liquid velocity and 
flux can be increased by a factor $\alpha$, of order $10$--$100$ or even more. For a binary mixture 
of water and a cosolvent, with viscosities $\eta_w$ and $\eta_{\rm cs}$, respectively, 
$\alpha$ increases monotonically with inter-plate potential $V$ and average ion content, and is large if 
the ratio $\eta_{\rm cs}/\eta_w$ is large.

\end{abstract}

\maketitle

\section{Introduction}

Confinement of liquids by solid surfaces is common in everyday life and in industrial applications. The 
relative movement between liquids and the surfaces that confine them affects wear and tear, 
lubrication, and energy consumption \citep{lugt_tribol_trans_2009,granick_tribol_lett_1998}.
These occur whether the surfaces move past each other or if the liquids are pumped in pipes or 
channels by external forcing. The importance of these processes has led to intensive 
industrial and fundamental research \citep{harvey_tribol_intl_2016,klein_arms_1996}. Control of the 
lubrication properties of liquids is desirable in many circumstances. For example, switchable drag 
reduction could provide a method to increase or decrease the flow in channels or pipes 
\citep{reis_adv_mat_2014}. Electrorheological fluids are particle suspensions that respond to the 
application of an external electric field \citep{hao_adv_mat_2001}. The long-range  interactions 
between the induced dipoles lead to particle chaining and a dramatic 
increase of the suspension's viscosity. In the viscoelectric effect, a liquid's viscosity changes due the 
coupling of an electric field with molecular dipoles \citep{dodd_nature_1939,dodd_prslsa_1946}.

Room-temperature ionic liquids attracted considerable interest in recent years as thin-film wear 
protectors and in energy applications \citep{licht_fpe_2004}. The arrangement of the molecules near a 
charged surface is not trivial due to screening of the field, packing in layers, steric effects, and size 
asymmetry between anions and cations 
\citep{urbakh_naturemat_2022,atkin_prl_2012,urbakh_nanolett_2023}.
The effective viscosity of a film can be measured with an atomic force microscope or surface force 
balance, and usually it increases with the charge of the confining surface  
\citep{urbakh_acs_anm_2020,urbakh_acsnano_2017}.

Recently, we found a new effect that we call ``electrolubrication'', whereby the effective viscosity of 
liquid mixtures can be controlled by electric fields \citep{tsori_pof_2023}. Here, a field transverse to the 
confining surfaces couples differently to the mixture's constituents, and screening by existing ions 
leads to {\it layering} of the mixture. The different viscosities of the pure constituents 
facilitate larger shear, which changes the flow. When the more polar liquid is the less 
viscous one, lubrication at surfaces is enhanced. In that work, we focused on lubrication between 
moving surfaces. The unperturbed velocity gradient (without imposed potential) and the resulting 
stress were constant . The maximal unperturbed velocity was at the moving surface, where the more 
polar liquid adsorbs. Here, we extend the theory to flows in channels between stationary walls, where 
the stress is not constant and the maximal unperturbed velocity is where less polar liquid is. The 
velocity vanishes at the boundary, where the polar liquid adsorbs.

\section{Model}

Consider a binary mixture of two liquids. The more polar liquid may be water, and the less polar one is 
a partially miscible cosolvent. The volume fraction, permittivity and viscosity of the water are 
$\phi$, $\veps_w$ and $\eta_w$, respectively. The same quantities for the cosolvent are denoted by 
$\phi_{\rm cs}$, $\veps_{\rm cs}$ and $\eta_{\rm cs}$. The mixture is coupled to a reservoir containing 
dissolved anions and cations whose number density is $n_0$. The water and cosolvent have partial 
miscibility given by their coexistence (binodal) curve. It is assumed that the bulk water  
composition, $\phi=\phi_0$ and temperature correspond to a homogeneous (mixed) state. While the 
theory is general, in some curves, as an example for a liquid pair we use the properties of water and 
glycerol. 
\begin{figure}
\includegraphics[width=0.48\textwidth,bb=1 560 540 725,clip]{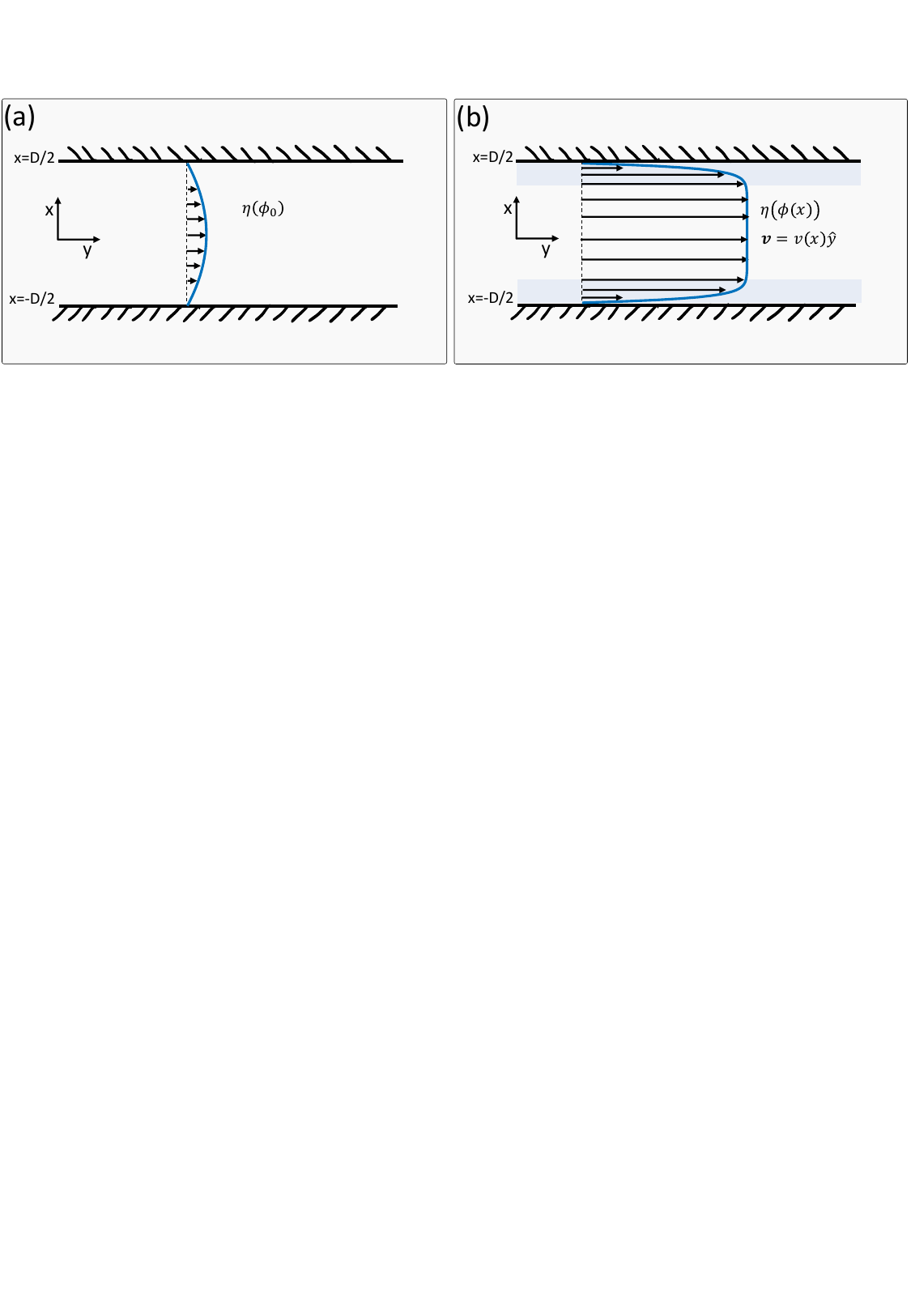}
\caption{Schematic illustration of the channel. Two flat and smooth walls parallel to $y$-$z$ plane are 
separated by a distance $D$ in the $x$-direction. The confined mixture is flowing along $y$. In 
steady state, the velocity depends on $x$: ${\bf v}=v(x)\hat{y}$. (a) When the mixture is homogeneous, 
its viscosity is $\eta(\phi_0)$ and the flow is the classic parabolic profile. (b) When a potential is applied 
across the walls, the mixture phase separates and two regions rich in the more polar solvent appear 
near the walls (faint blue shade). When the more polar solvent is the less viscous one, the lubrication 
layers modify the flow profile. The strong shear near the surfaces facilitates a large flux. 
}
\label{fig_illus}
\end{figure}

The mixture is pumped into a gap between two parallel and smooth surfaces, and flows parallel to the 
$y$-direction; see Fig. \ref{fig_illus}. 
The pressure gradient along the gap can result from a pump, gravitational force, etc. The length of 
the channel is $L$, and its walls are at $x=\pm D/2$. An electrostatic potential is 
applied across the channel in the $x$-direction. The resultant electric field modifies the mixture's 
composition and leads to composition gradients perpendicular to the main flow direction. 
We assume long channels where edge effects can be ignored. The system is then effectively 
one-dimensional, and all quantities depend on the $x$-coordinate only. In addition, we focus 
on steady-state conditions.

We start by formulating the flow behaviour based on a generic mixture free energy density $f$, and 
only later choose a specific model. The governing equations for the flow velocity ${\bf v}$, mixture 
composition $\phi$, ion densities $n^\pm$, and electrostatic potential $\psi$ are \citep{samin_prl_2017}
\bea
\nabla\cdot {\bf v}&=&0,\label{gov_eqns1}\\
\frac{\partial\phi}{\partial t}+{\bf v}\cdot\nabla\phi&=&\nabla\cdot\left(D_\phi\nabla\frac{\delta 
f}{\delta\phi}\right),
\label{gov_eqns2}\\
\rho\left(\frac{\partial{\bf v}}{\partial t}+({\bf v}\cdot \nabla){\bf v}\right)&=&-\nabla 
p+\nabla\left(\eta\nabla {\bf v}+\eta(\nabla{\bf v})^T\right)\nn\\
&-&\phi\nabla\frac{\partial f}{\partial\phi}+{\bf f}_{\rm elec},\label{gov_eqns3}\\
\nabla\left(\veps_0\veps\nabla\psi\right)&=&e\left(n^--n^+\right),\label{gov_eqns4}\\
\frac{\partial n^\pm}{\partial t}+\nabla\cdot(n^\pm{\bf v})&=&\nabla\cdot\left(D_i\nabla\frac{\delta 
f}{\delta 
n^\pm}\right).\label{gov_eqns5}
\eea

Equation (\ref{gov_eqns1}) is the condition of incompressibility, and Eq. (\ref{gov_eqns2}) is a 
diffusion-advection equation, with diffusion coefficient $D_\phi$. Equation (\ref{gov_eqns4}) 
is the Poisson equation, with $\psi$ the electrostatic potential, $e$ the electron's 
charge, and $\veps_0$ and $\veps(\phi)$ the vacuum and relative permittivities, respectively.
Equation (\ref{gov_eqns5}) is the Poisson-Nernst-Planck equation, with ion 
diffusion constant $D_i$ common to both ionic species. We use a modified Poisson-Boltzmann 
framework that takes the finite volume of the ions into account, thus
\bea
\phi+\phi_{\rm cs}+v_0n^++v_0n^-=1,
\eea
where $v_0$ is a volume common to all molecules -- water, cosolvent and ions. 
Equation (\ref{gov_eqns3}) is a Navier-Stokes equation with a force due to composition gradients, and a 
term ${\bf f}_{\rm elec}$, derivable from the Maxwell stress tensor, given by ${\bf f}_{\rm 
elec}=(1/2)\nabla\left(\veps_0 E^2 \phi\partial \veps/\partial \phi\right)_T-(1/2)\veps_0E^2\nabla\veps+
e(n^+-n^-){\bf E}$, with the electric field ${\bf E}=-\nabla\psi$. The composition-dependent 
permittivity is taken to be
\bea
\veps(\phi)=\veps_{\rm cs}\phi_{\rm cs}+\veps_w(1-\phi_{\rm cs}),
\eea
where $\veps_w$ and $\veps_{\rm cs}$ are the water and co-solvent permittivities, respectively.

The dependence of viscosity on composition is essential for electrolubrication. We assume that the 
ions' contribution to the viscosity is the same as that of water, and use the following linear constitutive 
relation:
\bea\label{eta_phi}
\eta(\phi)=\eta_{\rm cs}\phi_{\rm cs}+\eta_w(1-\phi_{\rm cs}).
\eea

The assumption of an infinitely long channel means that the flow velocity is parallel to the channel axis: 
${\bf v}=v(x)\hat{y}$. In steady state, all time derivative vanish. Since there is no flux of mixture or ions 
from the walls, we arrive at a remarkable simplification to (\ref{gov_eqns1})--(\ref{gov_eqns5}) --
both the composition profile $\phi(x)$ and ionic densities extremize the free energy: 
$\delta f/\delta\phi=0$ and $\delta f/\delta n^\pm=0$. As in equilibrium, they do not depend on 
$D_\phi$ and $D_i$.

Once $\phi(x)$ and $n^\pm(x)$ are found, one needs to solve the Navier-Stokes equation 
\bea\label{navier_stokes_simpl}
0=-\partial_y p_0+\partial_x\left(\eta\left(\phi(x)\right)\partial_xv(x)\right)
\eea
subject to the two boundary conditions $v(x=\pm D/2)=0$.

\subsection{Free energy} 

To be specific, we use below the free energy model 
\bea
\label{Omega}
\Omega&=&\int_{-D/2}^{D/2}\left(f-\lambda^+n^+-\lambda^-n^--\mu\phi/v_0 \right){\rm d}x+f_s,\nn\\
f&=&\frac12 c^2\left(\nabla \phi\right)^2+f_{\rm m}+f_{\rm e}+f_{\rm i}.
\eea
The channel walls are taken to be symmetric, hence the surface energy $f_s$ is taken as 
$f_s=\Delta\gamma(\phi(-D/2)+\phi(D/2))$, where $\Delta \gamma$ is the difference between the 
surface energies of water and cosolvent. The chemical potentials of the cations and anions are 
$\lambda^\pm$, respectively, and that of the mixture is $\mu$. The square-gradient term accounts for 
the energetic cost of composition inhomogeneities, where $c$ is a constant. 

The free energy density of mixing is
\begin{align}
\label{f_mix}
f_{\rm m}&=\frac{k_BT}{v_0}\left[
\phi\log(\phi)+\phi_{\rm cs}\log(\phi_{\rm cs})+\chi\phi \phi_{\rm cs}\right],
\end{align}
where $k_B$ is the Boltzmann constant, $T$ is the absolute temperature, and $\chi\sim1/T$ is the Flory 
interaction parameter. Equation (\ref{f_mix}) leads to an upper critical solution temperature type 
phase diagram. In the $\phi-\chi$ plane, a homogeneous phase is stable above the binodal curve 
$\phi_b(\chi)$, whereas below it the mixture separates to water-rich and water-poor phases, with 
compositions given by $\phi_b(\chi)$. The two phases become indistinguishable at the critical 
point $(\phi_c,\chi_c)=(1/2,2)$.

The electrostatic energy density $f_{\rm e}$ is given by
\begin{align}
\label{f_es}
f_{\rm e}=-\frac{1}{2}\veps_0\veps(\phi)(\nabla \psi)^2+e(n^+-n^-)\psi.
\end{align}
The free energy density of the ions , $f_{\rm i}$, is modelled as
\begin{eqnarray}\label{f_ions}
f_{\rm i}&=&k_BT\bigl[n^+(\log \left(v_0n^+)-1\right)+n^-\left(\log (v_0n^-) -1\right)\nn\\
&-&\phi(\Delta u^+n^+ + \Delta u^-n^-)\bigr].
\end{eqnarray}
The logarithmic terms account for the ions' entropy; the terms proportional to $\phi$ 
model specific chemical short-range interactions between ions and 
solvents. The parameters $\Delta u^\pm$ measure the preference of ions towards a water 
environment and are of order $\sim 1-10$ \citep{marcus_1988,tsori_jcp_2012,tsori_jcp_2013}. In the 
following, we deal with ions that are preferentially hydrophilic, and take them as equal: $\Delta 
u^\pm=\Delta u$.

\subsection{Free energy minimization} 

We use the following dimensionless variables: 
$\tl{x}=x/\lambda_D$, $\lambda_D^2=\veps_0\veps(\phi_0)k_BT/(2n_0e^2)$,
$\tl{\psi}=e\psi/k_BT$, $\tl{V}=eV/k_BT$, $\tl{n}^\pm=v_0n^\pm$, $\tl{n}_0=v_0n_0$,
$\tl{c}^2=c^2v_0/(\lambda_D^2k_BT)$, $\tl{\mu}=\mu/k_BT$,
$\tl{f}_m=v_0f_m/k_BT$ and $\Delta\tl{\gamma}=\Delta\gamma\lambda_D/c^2$. Here, 
$n_0$ and $\phi_0$ are the bulk ion density and mixture composition, respectively, $\lambda_D$ is the 
Debye length, and $V$ is the potential drop across the channel. Using these definitions, one can 
extremize the energy to find the profiles $\phi(x)$ and $\psi(x)$ as 
\bea\label{min_phi}
-\tl{c}^2\tl{\nabla}^2\phi+\frac{\partial 
\tl{f}_m}{\partial\phi}-\Delta u^+\tl{n}^+-\Delta u^- 
\tl{n}^-\nn\\
-\frac12\frac{2\tl{n}_0}{\veps(\phi_0)}\frac{\partial\veps}
{\partial\phi}(\tl{\nabla} \tl{\psi} )^2
-\tl{\mu}=0,\\
\tl{\nabla} \cdot (\veps(\phi)\tl{\nabla}
\tl{\psi})=\frac{\veps(\phi_0)}{2\tl{n}_0}(\tl{n}^--\tl{n}^+).&&\label{poisson}
\eea
The first equation expresses the variation with respect to $\phi$, i.e. $\delta f/\delta\phi=0$, and the 
second equation is Poisson's equation. The boundary conditions for $\phi$ and $\tl{\psi}$ are
\bea
\phi'(\tl{x}=\pm 
\tl{D}/2)&=&\mp\frac{\Delta\tl{\gamma}}{\tl{c}^2},\nn\\
\tl{\psi}(\tl{x}=\pm \tl{D}/2)&=&\pm \frac12 \tl{V}.
\eea

The ions obey a modified Boltzmann distribution \citep{andelman_prl_1997,tsori_cisc_2016}:
\bea\label{mod_pb}
\tl{n}^\pm&=&\frac{P^\pm(1-\phi)}{1+P^++P^-},\\
P^\pm &=&\frac{\tl{n}_0}{1-\phi_0-2\tl{n}_0} 
\exp\left[(\phi-\phi_0)(\chi+\Delta u^\pm)\mp\tl{\psi}\right].\nn
\eea

In general, the electric double layer created at the walls leads to adsorption of the polar solvent due to  
the dielectrophoretic force and preferential solvation. The thickness of the resulting lubrication layer is 
determined by both the Debye length and the correlation length of the mixture $\xi$, and thus can be 
large at temperatures close to the critical point \citep{tsori_pnas_2007,tsori_jcp_2013}. 

\subsection{Velocity profile}

Once the the composition profile $\phi(x)$ is known from (\ref{min_phi}), 
(\ref{poisson}) and (\ref{mod_pb}), $v(x)$ can be found in the following way from 
(\ref{navier_stokes_simpl}). We write the constant pressure drop along the channel 
as $-\partial_yp_0\equiv -\Delta p/L$, where $\Delta p$ is the pressure drop over a channel of length 
$L$. Using the symmetry of the problem with respect to the reflection $x\to -x$, one arrives at the 
solution for the flow profile $v(x)$ given by 
\bea\label{v_ef}
v_{\rm ef}(x)=\frac{\Delta p}{L}\int_{-D/2}^x\frac{x'}{\eta(\phi(x'))}dx'.
\eea
The no-flow condition at the walls, $v(x=\pm D/2)=0$, is satisfied. The total flux is given by the integral
\bea\label{flux_ef}
Q_{\rm ef}=\int_{-D/2}^{D/2} v(x)dx.
\eea

Recall that when the mixture is homogeneous with composition $\phi_0$ and viscosity 
$\eta_0=\eta(\phi_0)$, $v(x)$ is the classical parabolic profile
\bea\label{v_mixed}
v_{\rm m}(x)=\frac12\frac{\Delta p}{\eta_0 L}\left(x^2-\left(\frac{D}{2}\right)^2\right).
\eea
The flux of this mixed state is
\bea\label{flux_mixed}
Q_{\rm m}=-\frac{1}{12}\frac{\Delta p}{\eta_0 L}D^3.
\eea

The ``flow amplification factor'' $\alpha$ is defined as the ratio between the fluxes with and without 
electrostatic potential:
\bea\label{alpha}
\alpha=\frac{Q_{\rm ef}}{Q_m},
\eea
with the integrated fluxes $Q_{\rm ef}$ and $Q_m$ taken from (\ref{flux_ef}) and 
(\ref{flux_mixed}), respectively. 

In the next section we solve numerically the profiles $\phi(x)$ and $\psi(x)$, evaluate the velocity and 
flux from (\ref{v_ef}) and (\ref{flux_ef}) for the state with an electric field, and compare them with 
the same quantities without the field.

\section{Results}

We solve (\ref{min_phi}) and (\ref{poisson}) numerically as diffusion equations with pseudo-time 
until the solutions do not change within the desired error. Spatial 
derivatives were discretized using standard schemes and the time integration was done using the
Runge-Kutta algorithm. The results were substituted in the equations to verify their validity. For 
example, in (\ref{poisson}), the right- and left-hand sides were verified to be equal 
within a relative error of $10^{-3}$. We used channels with scaled width $D=8\lambda_D$ and several 
values of cross-channel potential $\tl{V}$.
\begin{figure}
\includegraphics[width=0.48\textwidth,bb=0 0 805 295,clip]{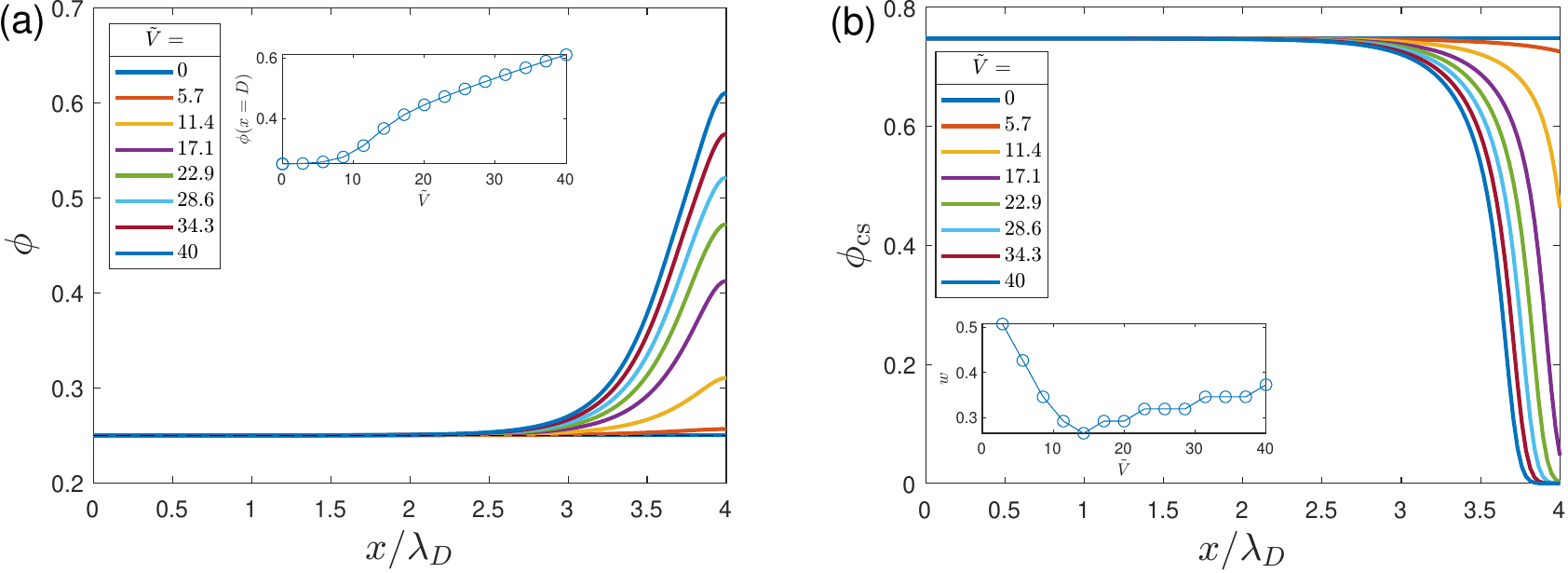}
\caption{(a) Water ($\phi$)  profiles for varying values of scaled potential $\tl{V}$ (see legend). The total 
channel width is $D=8\lambda_D$, and the walls are located at $x/\lambda_D=\pm 4$. The inset shows 
the surface value $\phi(x=D)$ versus $\tl{V}$. (b) Cosolvent ($\phi_{\rm cs}$) profiles for the same 
potentials. The inset is the width of the wetting layer close to the wall $w$, defined as the distance 
from the walls, where $\phi$ is the average between $\phi_0$ and $\phi(x=D)$. In all figures, unless 
stated otherwise, we used $\phi_0=0.25$, $\chi=1$, $\Delta u=2$, $\veps_w=80$, $\veps_{\rm cs}=45$, 
$\tl{c}=0.4$, $\eta_{\rm cs}/\eta_w=1412$, $\Delta\tl{\gamma}=0$ and $\tl{n}_0=0.001$. 
}
\label{fig_phi_profiles_var_Vt}
\end{figure}

Figure \ref{fig_phi_profiles_var_Vt}(a) shows the water composition, and figure 
\ref{fig_phi_profiles_var_Vt}(b) the cosolvent composition, across the 
channel as a function of increasing potential. When $\tl{V}=0$, $\phi$ equals the bulk composition 
$\phi_0=0.25$. As $\tl{V}$ increases, two boundary layers appear at the walls. The value of $\phi$ 
increases near the walls, while $\phi_{\rm cs}$ decreases. Their sum is not exactly unity since the 
dissolved ions have finite volumes. Note that $\tl{V}=40$ corresponds to a physical potential of $1$ V. 
Clearly, at the walls, water is enriched, therefore the viscosity there is greatly reduced locally as 
compared to the centre of the channel ($x=0$). The inset in figure \ref{fig_phi_profiles_var_Vt}(b) shows 
the width of the water layer near the walls, $w$, which depends on the electrostatic screening length 
and on the mixture's correlation length. The non-monotonic behaviour of $w$ versus $\tl{V}$ is a result 
of the simultaneous nonlinear decrease of screening and the stronger adsorption of water with 
increasing $\tl{V}$.
\begin{figure}
\includegraphics[width=0.48\textwidth,bb=0 0 815 295,clip]{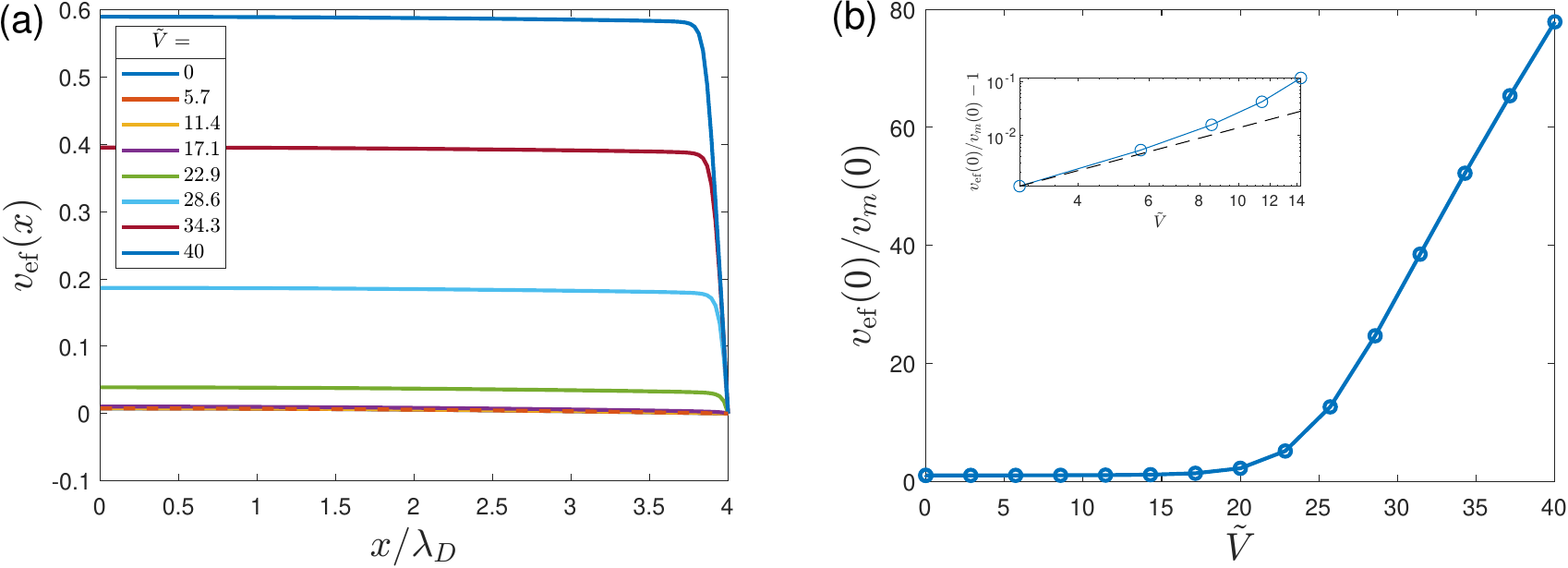}
\caption{(a) Velocity profile $v(x)$ across the channel from (\ref{v_ef}) for varying values of scaled 
potential $\tl{V}$ (see legend). When $\tl{V}=0$, the mixture is homogeneous. The flow is then 
parabolic and given by (\ref{v_mixed}). It is not visible because its amplitude is too small. Here, 
$v_{\rm ef}$ is given in units of $\Delta p\lambda_D^2 /(L\eta_w)$. (b) The ratio of the mid-channel 
velocities with and without electric potential, $v_{\rm ef}(0)/v_m(0)$, for varying potentials. The inset 
is a log-log plot indicating that $v_{\rm ef}(0)/v_m(0)-1\sim \tl{V}^2$ (slope of dashed line is $2$).
}
\label{fig_vel_profiles_var_Vt}
\end{figure}

Figure \ref{fig_vel_profiles_var_Vt} (a) shows the resultant velocity profiles $v_{\rm ef}(x)$ calculated 
from (\ref{v_ef}) for the same values of $\tl{V}$. The curves are characterized by a thin boundary 
layer at the walls, where the gradient of $v_{\rm ef}(x)$ is large, and a large region, far from the walls, 
where the velocity is high and approximately constant. As $\tl{V}$ increases, the maximal speed in the 
centre increases, and the gradient $v'_{\rm ef}$ near the walls becomes even larger. Using the 
numerical values $\Delta p=1$ atm, $L=0.01$ m and $\lambda_D=10$ nm, 
we estimate $v_{\rm ef}$ to be of the order of $1$ $\mu$m/sec for the largest potential, $\tl{V}=40$. 
This is almost $80$ times larger than the velocity in the absence of potential.

In figure \ref{fig_vel_profiles_var_Vt}(b), we plot the ratio of the velocities with and without a field,  
$v_{\rm ef}/v_m$. Both are evaluated at their maximum, i.e. at $x=0$. and $v_m(x)$ is the parabolic 
profile in (\ref{v_mixed}). When $\tl{V}=0$, the two are equal. The ratio increases with $\tl{V}$ to high 
values -- at a physical potential of just $1$ V, $v_{\rm ef}(0)/v_m(0)$ is near $80$.

\begin{figure}
\begin{minipage}{0.5\textwidth}
\includegraphics[width=\textwidth,bb=1 1 435 245,clip]{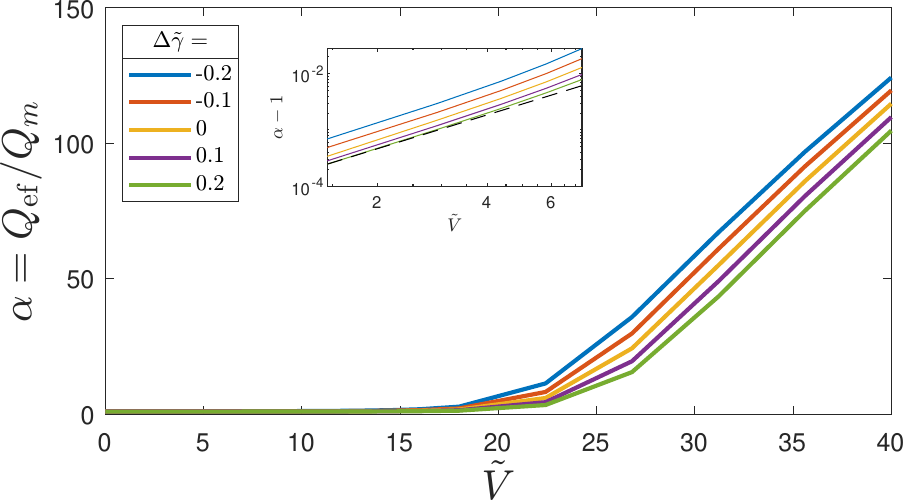}
\end{minipage}
\begin{minipage}{0.5\textwidth}
\caption{Flow amplification factor $\alpha$ versus $\tl{V}$ for different values of 
$\Delta\tl{\gamma}$. Here, $\alpha$ in (\ref{alpha}) is defined as the ratio 
of the total channel flux with and without electric potential, with $Q_{\rm ef}$ and $Q_m$ taken from 
(\ref{flux_ef}) and (\ref{flux_mixed}), and $Q_m$ is the flux of the classical parabolic profile, in the 
absence of potential. The curves differ by the value of $\Delta \tl{\gamma}$ (see legend). As the walls 
become more hydrophilic (decreasing value of $\Delta\tl{\gamma}$), the flux increases relative to 
$Q_m$. The inset is a log-log plot indicating that $\alpha-1\sim\tl{V}^2$ (slope of dashed line is $2$).
}
\label{fig_alpha_var_Dgt}
\end{minipage}
\end{figure}

In the calculations thus far, the walls were neither hydrophilic nor hydrophobic, 
$\Delta\tl{\gamma}=0$. It is interesting to study how the flow is affected by the relative
hydrophilicity of the surfaces by allowing $\Delta\tl{\gamma}$ to vary. Figure \ref{fig_alpha_var_Dgt} 
shows the ``flow amplification factor'' $\alpha$ from (\ref{alpha}). As expected, the general trend is 
that $\alpha(\tl{V})$ increases with $\tl{V}$. Each curve corresponds to a different value of 
$\Delta\tl{\gamma}$. While all curves reach large flow amplification at $\tl{V}=40$ ($\alpha\sim 100$ 
for all curves), it is clear that $\alpha$ becomes larger monotonically as the surface hydrophilicity 
increases. 

\begin{figure}
\begin{minipage}{0.5\textwidth}
\includegraphics[width=\textwidth,bb=1 1 435 245,clip]{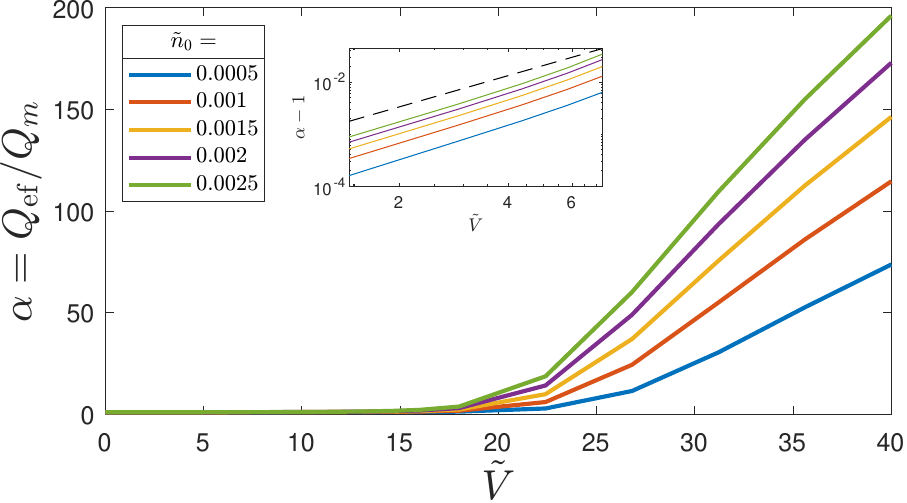}
\end{minipage}
\begin{minipage}{0.5\textwidth}
\caption{Flow amplification factor $\alpha$ versus $\tl{V}$ for different salt contents; the legend
indicates the value of $\tl{n}_0$. As salt is added, the walls adsorb more water, and the flux increases 
relative to $Q_m$, with $\Delta\tl{\gamma}=0$. In all curves, $D=8\lambda_D$, with $\lambda_D$ 
depending on $\tl{n}_0$. The inset is a log-log plot indicating that $\alpha-1\sim\tl{V}^2$ (slope of 
dashed line is $2$).
}
\label{fig_alpha_var_n0t}
\end{minipage}
\end{figure}
In figure \ref{fig_alpha_var_n0t}, we look at the dependence of electrolubrication on the salt content. 
We calculated $Q_{\rm ef}$ versus $\tl{V}$ with a given amount of salt $\tl{n}_0$, and then repeated 
with more salt. While $\alpha$ increases with $\tl{V}$ and reaches high values at $\tl{V}=40$, it also 
increases monotonically as salt is added. 

\begin{figure}
\includegraphics[width=0.48\textwidth,bb=0 0 827 295,clip]{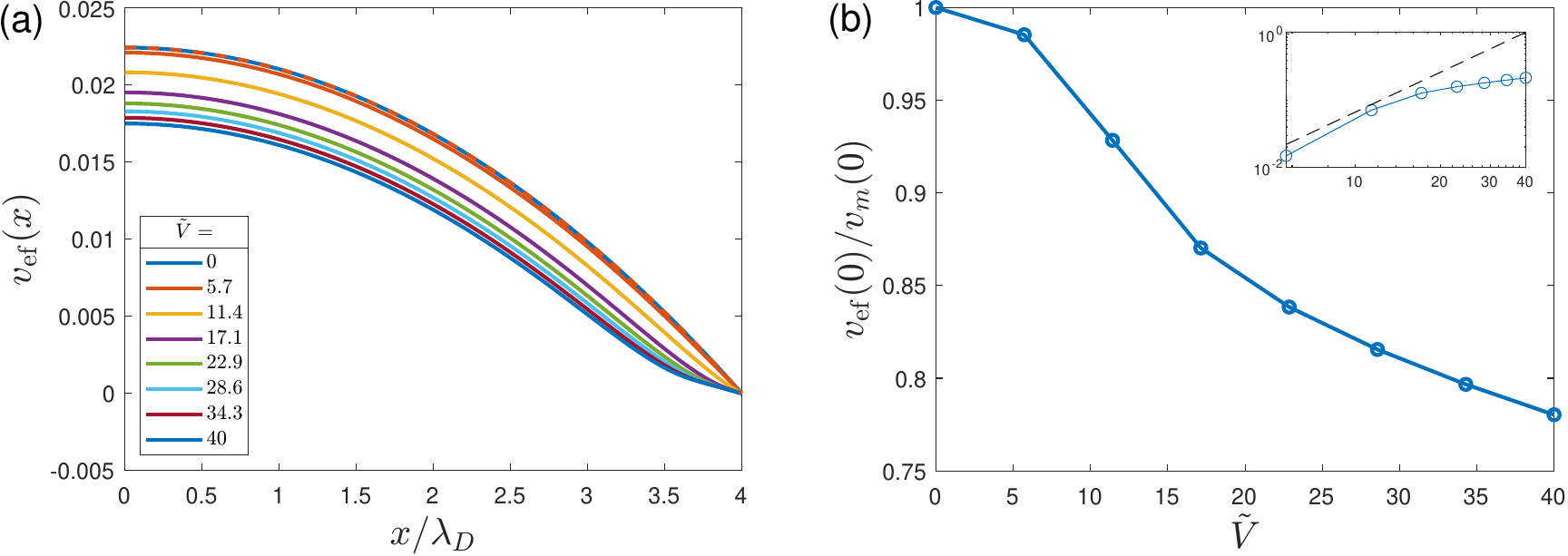}
\caption{The same as in Fig. \ref{fig_vel_profiles_var_Vt} but now the polar liquid is the more viscous 
one, i.e., assuming $\eta_w/\eta_{\rm cs}=1412$. Dashed line in (a) is the classical parabolic profile 
$v_m(x)$. The limit $\tl{V}\to\infty$ describe complete separation between the liquids, where 
their velocities given by Eqs. (\ref{vel_two_regions}). When $\eta_w\gg\eta_{\rm cs}$, and using 
$\phi_0=2w/D$, one finds that the curve in (b) tends to $v_{\rm ef}(0)/v_m(0)\approx (1-\phi_0)^2=0.56$ 
in the large potential limit when $\phi_0=0.25$. Inset is a log-log plot of $1-v_{\rm ef}(0)/v_m(0)$ vs 
$\tl{V}$. The slope of the dashed line is $2$.
}
\label{fig_vel_profiles_var_Vt_inverse}
\end{figure}
Until now, the polar liquid (water) was considered to be less viscous than the non-polar one 
(cosolvent), thereby facilitating the two lubrication layers when potential is applied across the walls. 
It is tempting to think of the opposite case, where the polar liquid is more viscous. If this liquid is 
very viscous, then could the two layers at the walls act as ``pipe cloggers'', effectively reducing the 
channel width $D$ (or pipe diameter in the case of a circular pipe)?

Figure \ref{fig_vel_profiles_var_Vt_inverse} examines this situation. All parameter values are the same 
as in figures \ref{fig_phi_profiles_var_Vt} and \ref{fig_vel_profiles_var_Vt}, except that the viscosities are 
interchanged: $\eta_w/\eta_{\rm cs}=1412$. In figure \ref{fig_vel_profiles_var_Vt_inverse}(a), $v_{\rm 
ef}$ is shown for varying values of $\tl{V}$. The flow velocity is slower at any point $x$, and, in 
particular, in the centre. Figure \ref{fig_vel_profiles_var_Vt_inverse}(b) quantifies the reduction of the 
flow velocity in the center. At the maximum value, $\tl{V}=40$, the velocity is reduced by a modest 
$\sim 25\%$ from its no-field case. 
The inset shows that $1-v_{\rm ef}(0)/v_m(0)\sim \tl{V}^2$.  

The figures above show a dependence on $\tl{V}^2$ for small potentials. The small $\tl{V}$ limit can be 
obtained as follows: (\ref{min_phi}) is linearized around the bulk composition $\phi_0$ using the 
Debye-H\"{u}ckel solution of (\ref{poisson})). The homogeneous solution includes exponents with 
the mixture correlation length and the particular solution includes the Debye length. Next, the 
Navier-Stokes equation (\ref{navier_stokes_simpl}) is solved by 
assuming that $v(x)=v_0(x)+\delta v(x)$, where $v_0$ is the parabolic profile, and $\delta v$ is a small 
perturbation whose boundary values are $\delta v(x=\pm D/2)=0$. The dependence of $\delta v$ on 
$\phi$ is via (\ref{eta_phi}). The perturbation $\delta v$ and the corresponding flux are 
proportional to $\tl{V}^2$. As a result, $1-v_{\rm ef}(0)/v_m(0)\sim\tl{V}^2$.

\begin{figure}
\includegraphics[width=0.48\textwidth,bb=0 0 810 265,clip]{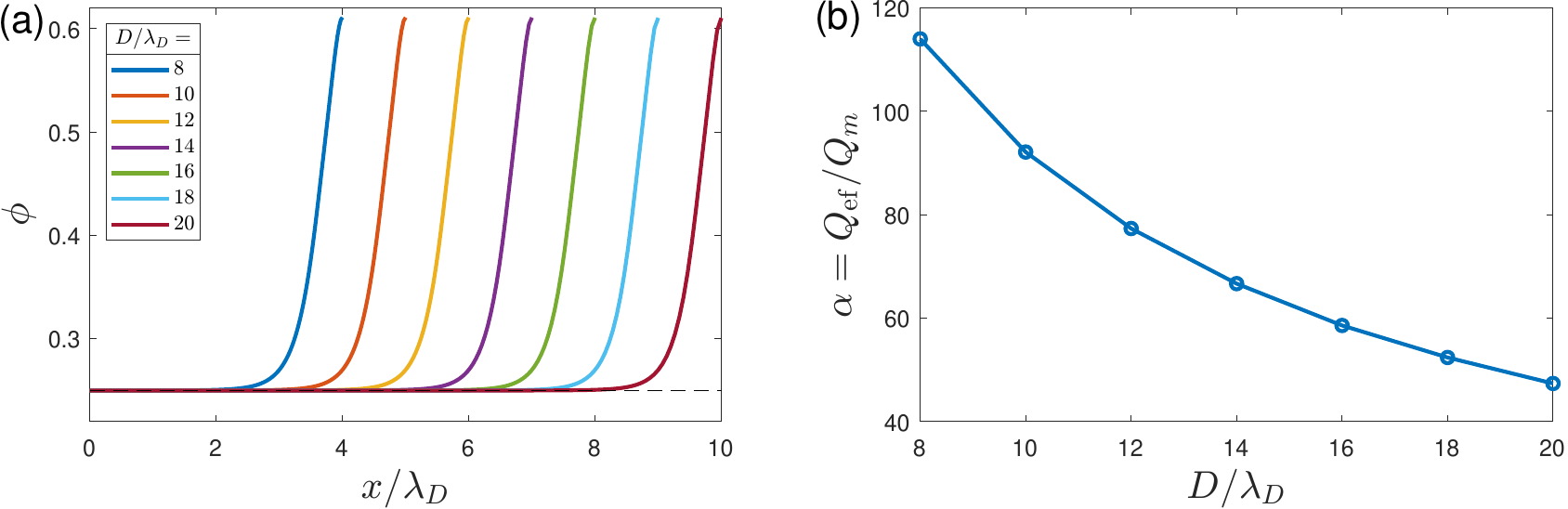}
\caption{(a) Composition profiles for different channel widths $D$ (see legend) at constant potential 
$\tl{V}=40$. Each curve has a different $x$-range: for example, the dark blue curve corresponds to 
$D=8\lambda_D$, and therefore the $x$-range is $-4\leq x/\lambda_D\leq 4$. All curves have a wetting 
layer of the less viscous liquid at $x\lesssim D$. Calculations assume a bulk value $\phi_0=0.25$ far 
from the walls. (b) Flow amplification ratio $\alpha$ versus channel width $D$. Here, $\alpha$ 
decreases with $D$ since the relative volume fractions of the wetting layers decreases with $D$. 
}
\label{fig_phi_profiles_var_D}
\end{figure}
How does $\alpha$ scale with the channel width $D$?
In Fig. \ref{fig_phi_profiles_var_D}, we varied $D$ at constant pressure gradient and potential. Figure 
\ref{fig_phi_profiles_var_D}(a) shows the profiles $\phi(x)$ versus $x$ for several values of $D$ 
(different colours; see legend). Note that each curve has a different $x$-range. All curves exhibit a 
wetting layer near the walls. These layers are similar to each other because the potential is the same. 
The calculations assume that far from the surfaces, the composition is fixed at $\phi_0=0.25$. Figure 
\ref{fig_phi_profiles_var_D}(b) shows $\alpha$ versus $D$.
At small channel widths, $\alpha$ is large since the volume fraction of the less viscous solvent (water) 
is high. As $D$ increases, the volume fraction of the wetting layers becomes smaller, and $\alpha$ 
decreases monotonically. In the limit $D\to\infty$, $\alpha$ tends to unity.

We now estimate $\alpha$ by using a simple analytical model, where two layers of pure water of 
thickness $w$ are at the walls, and pure cosolvent is in the centre. Hence $\phi_0=2w/D$ is the water 
volume fraction. From (\ref{v_ef}), one finds the velocity in the two regions to be
\bea\label{vel_two_regions}
v_{\rm cs}(x)&=&\frac12 \frac{\Delta p}{\eta_{\rm cs}L}x^2+c_1,~|x|\leq D/2-w,\nn\\
v_w(x)&=&\frac12 \frac{\Delta p}{\eta_wL}x^2+c_2,~|x|>D/2-w.
\eea
From the continuity of $v$ at $x=\pm (D/2-w)$, and the no-flow boundary conditions, it follows that 
$c_2=(-1/8)\Delta p D^2/(\eta_w L)$ and $c_1=(\Delta p/2L)(1/\eta_w-1/\eta_{\rm 
cs})(D/2-w)^2-(1/8)\Delta p D^2/(\eta_wL)$.

\begin{figure}
\begin{minipage}{0.5\textwidth}
\includegraphics[width=\textwidth,bb=0 0 430 250,clip]{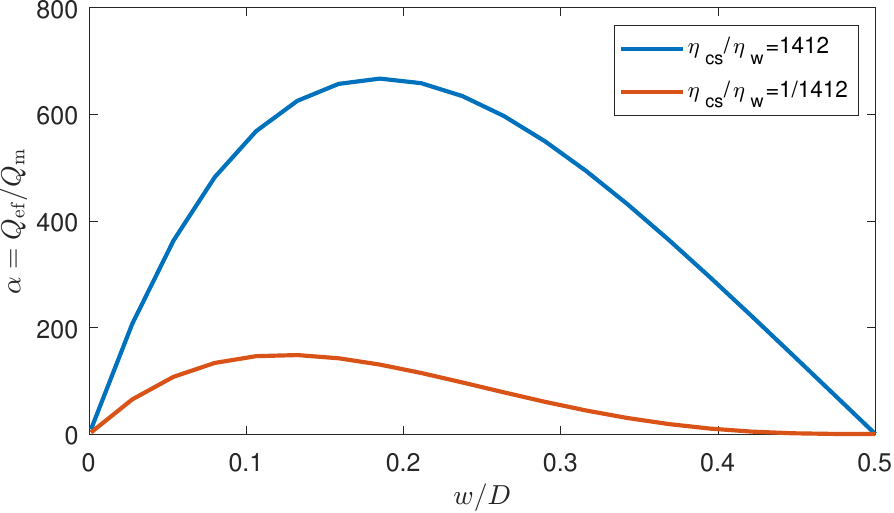}
\end{minipage}
\begin{minipage}{0.5\textwidth}
\caption{Flow amplification factor $\alpha$ as a function of model layer thickness $w$. This layer 
comprises the more polar liquid, and its viscosity is $\eta_w$. The viscosity of the non-polar liquid in the 
centre of the channel is $\eta_{\rm cs}$. The blue curve is where the polar layer is less viscous; the red 
curve is where the polar layer is more viscous.
}
\label{fig_alpha_var_w_analytical}
\end{minipage}
\end{figure}

The flux integrated over the channel width is
\bea
Q_{\rm ef}&=&2\int_0^{D/2-w}v_{\rm cs}dx+2\int_{D/2-w}^{D/2}v_wdx\\
&=&-\frac{1}{12}\frac{\Delta p}{L}\frac{2w\left(3D^2-6Dw+4w^2\right)\eta_{\rm 
cs}+\left(D-2w\right)^3\eta_w}{\eta_{\rm cs}\eta_w}.\nn
\eea
The flux of the mixed state can be obtained from (\ref{flux_mixed}), with the average viscosity given 
by $\eta_0=(2w/D)\eta_w+(1-2w/D)\eta_{\rm cs}$:
\bea
Q_m=-\frac{1}{12}\frac{\Delta p}{L}\frac{D^3}{\frac{2w}{D}\eta_w+(1-\frac{2w}{D})\eta_{\rm cs}}.
\eea
The ratio $\alpha=Q_{\rm ef}/Q_m$ increases monotonically with the viscosity ratio $\eta_{\rm 
cs}/\eta_w$ when $w/D$ is held constant. One can also hold the ratio $\eta_{\rm cs}/\eta_w$ constant 
and inspect $\alpha$ as a function of $w/D$. The small $w/D$ values correspond to the large $D$ 
range in Fig. \ref{fig_phi_profiles_var_D} (b). Surprisingly, as can be seen in Fig. 
\ref{fig_alpha_var_w_analytical}, $\alpha$ has a maximum at a finite lubrication layer thickness $w$. 
The blue curve shows $\alpha$ for the case where the less polar liquid is more viscous ($\eta_{\rm 
cs}/\eta_w=1412$). In the red curve, the viscosities are interchanged: the polar phase is more viscous, 
$\eta_w/\eta_{\rm cs}=1412$. The existence of the maximum is due to the assumption of a 
closed system, where $\eta_0$ depends on $w$, $D$, $\eta_{\rm cs}$ and $\eta_w$, which is different 
from a system coupled to a reservoir, where $\eta_0=\eta_{\rm cs}$.  

\section{Conclusion}

We investigate the electrolubrication of liquid mixtures flowing between two parallel plates. 
An electrostatic potential applied across the surfaces causes partial demixing of the liquids. The more 
polar liquid is adsorbed to the walls, due to screening of dissolved ions. The thickness of the lubrication 
layer can be large at temperatures close to the critical point. A similar phenomenon could be achieved, 
in principle, without ionic screening, but the geometry would have to allow for field gradients 
\citep{tsori_efips_2004}. In addition, the potentials required are higher. 

When the more polar liquid is less viscous than the non-polar one, shear stress falls mainly on 
the thin lubrication layers, and the flow profile is modified significantly from the classical parabolic one. 
The velocity and flux in the gap are then increased. The `flow amplification factor' $\alpha$, 
measuring the relative increase in flux, is of order $10-100$ or more for mixtures of water and glycerol. 
It increases monotonically with applied potential. Additionally, $\alpha$ depends on the viscosity ratio, 
ionic content, temperature and surface tension with the walls.

The flow profile in Fig. \ref{fig_illus} (b) is reminiscent of yield stress phenomena in Bingham fluids. 
Usually in yield stress phenomena, the fluid is homogeneous, and the yielding occurs when the 
pressure is larger than a critical value. Here, the fluid is made up from two liquids, and ``yielding'' is 
due to the field acting transverse to the flow while the pressure is constant. 

One possible application of 
`electrlubrication' is in microfluidic devices, whereby the flow of liquid mixtures flowing in narrow 
channels could be manipulated by transverse potentials. Electrical contacts are already commonly 
used to control the location and movement of droplets transport in channels. As we show in this paper, 
precise temperature regulation or exact solution composition are not required. Another possible 
application is in micro electromechanical systems, where small, solid, moving elements slide past each 
other. If the liquid embedding the moving parts is a mixture, then control of the elements' charge 
would allow us to modify the lubrication layer around the elements, and therefore the friction. In many 
cases, solid surfaces are charged (e.g. silica in aqueous water), and then it is important to asses the 
lubrication effect.

It would be interesting to lift the assumption of steady state and study e.g. a homogeneous 
mixture entering a constriction subject to strong electric field. The mixture's composition and velocity 
then evolve along the flow direction, and reach steady state only sufficiently far downstream. The 
understanding of such dynamical processes is not trivial and could be important for various 
applications of electrolubrication.

{\bf Acknowledgment} 
This work was supported by the Israel Science Foundation (ISF) Grant No. 
274/19.


%
%


%


\end{document}